\documentclass{aa}
\usepackage{graphicx}
\usepackage{bm,natbib,url,wasysym}
\usepackage[varg]{txfonts}
\usepackage{xcolor}
\usepackage{amsmath}
\usepackage[colorlinks,allcolors=blue]{hyperref}
\bibpunct{(}{)}{;}{a}{}{,} 
\graphicspath{{./figs/}{./png/}}

\newcommand{\EQ}{\begin{equation}}
\newcommand{\EN}{\end{equation}}
\newcommand{\EQA}{\begin{eqnarray}}
\newcommand{\ENA}{\end{eqnarray}}

\newcommand{\Eq}[1]{Equation~(\ref{#1})}

\newcommand{\Sec}[1]{Section~\ref{#1}}

\newcommand{\Fig}[1]{Fig.~\ref{#1}}

\newcommand{\Tab}[1]{Table~\ref{#1}}

{}
{}
{}

{}
{}
{}
{}
{}
{}
{}
{}
{}
\newcommand{\meanBB}{\overline{\mbox{\boldmath $B$}}{}}{}
{}
{}
{}
{}
{}
{}
{}
{}

{}
{}
{}

{}
{}
{}
{}

{}

{}
{}

\newcommand{\uu}{\mbox{\boldmath $u$} {}}

\newcommand{\BB}{\mbox{\boldmath $B$} {}}

\newcommand{\JJ}{\mbox{\boldmath $J$} {}}
\newcommand{\SSS}{\mbox{\boldmath $S$} {}}
\newcommand{\AAA}{\mbox{\boldmath $A$} {}}

\newcommand{\nab}{\mbox{\boldmath $\nabla$} {}}

\begin{document}

\titlerunning{Magnetic helicity enhances coronal heating and X-ray emission}
\authorrunning{Warnecke \& Peter}

\title{On the influence of magnetic helicity on X-rays emission of solar and stellar coronae}
\author{J. Warnecke\inst{1} \and H. Peter\inst{1}}
\institute{Max-Planck-Institut für Sonnensystemforschung,
  Justus-von-Liebig-Weg 3, D-37077 G\"ottingen, Germany\\
\email{warnecke@mps.mpg.de}\label{inst1} 
}

\date{\today,~ $ $Revision: 1.5 $ $}
\abstract{
Observation of solar-like stars show a clear relation between X-ray
emission and their rotation. Higher stellar rotation can lead to a
larger magnetic helicity production in stars.
}{We aim to understand the relation between magnetic helicity on the
  surface of a star to their coronal X-ray emission.  
}{We use 3D MHD simulations to model the corona of the solar-like
  stars. We take an observed magnetogram as in photospheric activity
  input, and inject different values of magnetic helicity. We
  use synthesis emission to calculate the X-ray emission flux of each
  simulation and investigate how this scales with injected magnetic helicity.
}{We find that for larger injected magnetic helicities an increase in
  temperature and an increase in X-ray emission. The X-ray emission
  scaled cubicly with the injected helicity. We can related this to
  increase of horizontal magnetic field and therefore higher Poynting
  flux at the coronal base.
}{
Using typical scaling of magnetic helicity production with stellar
rotation, we can explain the increase of X-ray
emission with rotation only by an increase of magnetic helicity at the
surface of a star.
}

\maketitle

\section{Introduction}

Solar-like stars show strong dependence of magnetic activity with rotation.
This is most clearly seen in the enhancement of X-ray emission
\citep[e.g.][]{PMMSBV03,VGJDPMFBC14,RSP14,WD16} for increasing stellar
rotation until a plateau in activity is reached.
Understanding the overall relation of rotation to emission is far
from trivial, because several different processes are involved: the
generation and surface appearances of magnetic field and the magnetic
heating in the stellar atmosphere.

Rotation is a key ingredient for the magnetic field generation below
the stellar surface via a dynamo process \citep{BS05}. An increase of
rotation is believed to lead to a more efficient dynamo producing
larger magnetic fields. They appear at the surface in form of larger
spots and heating the stellar coronae to higher temperatures, leading
to higher X-ray emission. As found in
solar-like stars, the integrated X-ray emission can be mostly
explained via thermal radiation with a temperature to the power of
4.5$\pm$0.3 \citep{guedel:2004}.
This scenario is supported observationally by a clear relation between stellar
rotation and the large-scale surface magnetic field
\citep{VGJDPMFBC14}. 
Also numerical simulations of stellar dynamos indicate a clear trend in this direction
\citep[e.g.][]{SPRD14,VWKKOCLB18,W18,ABT19,WKK19}. Observations show further
evidence for increasing coronal X-ray emission for larger magnetic
surface flux \citep[e.g.][]{PFALJ03,VGJDPMFBC14}. However, the
detailed process, how larger magnetic surface flux lead to higher
temperatures in the stellar corona is not fully understood as in
particular the small-scale granular motion play an important part in
generating the necessary upward directed Poynting flux.

Beside leading to larger magnetic field via the dynamo process,
stellar rotation can also influence the magnetic field topology
appearing at the stellar surface.
Below the surface, the interplay of convection and rotation is an
essential part in the magnetic field generation via the $\alpha$ effect
\citep{SKR66}. An $\alpha$ effect will produce helical magnetic field
with a preferred handedness on the large-scales and the opposite on the
small-scales \citep{BS05} as recently confirmed to be present in the
Sun \citep{SKBKLV18}. The helical nature of the magnetic field,
expressed in terms of the magnetic helicity, depends crucially on the value of
rotational influence on convection. Higher rotation lead to convective
motion carrying higher helicity and therefore generating more
helical magnetic fields \citep{KR80}.
Therefore, one would expect that stars more rapidly rotating than the Sun
shows more helical magnetic field at the stellar surface. This
rotational dependence in magnetic topology in turn
would also have an influence on the heating of the stellar coronae
and therefore their X-ray production.

Observational studies relating magnetic helicity and X-ray emission
has been limited to the work by \cite{MKYS05}, where they found that
the X-ray flux of solar active region scales with the injection of magnetic
helicity flux with a power of
1.5. However, they found a tighter correlation with the magnetic surface
flux of these regions, which is more likely to be the cause of the
X-ray variations.
In this paper, we want to investigate how an increase of magnetic
helicity at the surface changes the X-ray production of stellar
coronae in-depended on magnetic surface flux.

Magnetic helicity is a conserved quantity in ideal
magnetohydrodynamics and even in the non-ideal case it decays slower
than the magnetic energy \citep{M78}. 
Therefore, it is know the play an crucial role in the connecting the
rotational influence magnetic field dynamics below the surface to
the one above the surface.
It has been shown that magnetic helicity play a key role in the
triggering of eruptive events on the Sun, e.g. solar flare and solar
coronal mass ejections \citep[e.g.][]{NZZ03,Pariat+al:2017}. Furthermore,
there are indication that magnetic helicity also plays an important
role the heating in coronal loops \citep{WCBP17,BSB18}.
However, how the coronal heating scales with magnetic helicity have not been
studied so far.

To measure magnetic helicity on the solar surface is not trivial.
Magnetic helicity is defined as the volume integral of
$\AAA\cdot\BB$, involving not only the magnetic field $\BB$, but also
its vector potential $\AAA$. Hence, one needs to estimate/measure
$\AAA$ and to cope with the gauge dependency of $\AAA$.
Commonly, the current helicity, the dot product of magnetic field and
current density, is used as a proxy for magnetic
helicity, It is gauge invariant and its vertical contribution can
be measured from magnetograms \citep[e.g.][]{ZSPGSK10}.
Others invoke the gauge invariant relative helicity \citep{BF84} to
calculate the injection of helicity flux \citep[e.g.][]{CWQGSY01,NZZ03,MKYS05,V19} based on the
photospheric motions determined by local correlation tracking.
Another approach used the vector magnetograms of active region to
calculate the magnetic helicity spectrum. Because this method assumes
periodicity in horizontal direction is makes the obtained spectrum
gauge invariant. This method was used to determine the value of
magnetic helicities in several active regions \citep{ZBS14,ZBS16}. They found
10$^{4}$ up to 5$\times10^{5}$ G$^2$Mm for magnetic helicity density,
see also the followup study by \citep{GB19}.
In our paper, we include various values of the magnetic helicity
density at the photospheric surface and study with an established
model of the solar corona \citep{BP11,BP13,WP19}, how the helicity influences the heating and
X-ray production in the coronae.

The paper is structured in the following way. We first described the
basics of the numerical model including how we inject magnetic
helicity in the photosphere in \Sec{sec:model}. Then we discuss the magnetic helicity
evolution and distribution in \Sec{sec:hel}, its influence on X-ray
emission production in \Sec{sec:xray}, and how X-ray emission,
temperature an Poynting flux scales with helicity in
\Sec{sec:scale}. Before we conclude in \Sec{sec:con}, we also discuss
the relation between the magnetic helicity, X-rays and extreme UV
emission \Sec{sec:euv}.

\section{Model and setup}
\label{sec:model}

We model the solar and stellar corona in a Cartesian box ($x$,$y$,$z$) starting from
the photosphere ($z$=0) to the corona ($z$=80 Mm) with an horizontal
extent of 100 Mm. We use the {\sc Pencil Code}\footnote{\url{https://github.com/pencil-code/}} to solve the
equation of compressible resistive magnetohydrodynamics. 
This includes the induction equation for the vector
potential $\AAA$, which assures the solenoidality of the magnetic
field $\BB=\nab\times\AAA$, the momentum equation for the velocity
$\uu$, the continuity equation for $\rho$ and equation of state for an
ideal gas. The exact setup is described in detail in the work by
\cite{BP11,BP13}. The latest additions including the semi-relativistic
Boris correction to the Lorentz force \citep[e.g.][]{Gombosi02, CP18} and the non-Fourier description of the heat flux
evolution to speed up the calculation are presented in \cite{WB18}.
Key element of this model is a realistic description of the Spitzer heat conductivity, which
is along the magnetic field and depend strongly on the temperature
$T^{5/2}$.
The values of constant viscosity $\nu$ and magnetic diffusivity $\eta$
are chosen in such a way that their corresponding grid Reynolds numbers are around unity.
This value of $\nu$ is close the realistic Spitzer value in the solar
corona, but $\eta$ and therefore the magnetic Prandtl number are
several orders of magnitude different to realistic values.
To avoid strong artificial currents and hence large ohmic heating near the top
boundary we use an slightly larger value of all diffusivities and a
reduced Ohmic heating term near the top boundary.
We use periodic boundary conditions in the horizontal direction for
all quantities. At top boundary, we use vanishing values for the
velocity and heat flux with hydrostatic extrapolation for the density.
The magnetic field is following a potential field extrapolation at
both boundaries. At the bottom boundary, we prescribed the horizontal velocity using a
granulation driver, which mimics the photospheric velocities of the Sun
\citep{GN02,BP11}. Temperature and density are fixed at the bottom
boundary.
One important ingredient of this model is that we drive the simulations
by an observed magnetogram for the vertical magnetic field.
For this work, we use line-of-sight magnetic field from the active region
AR 11102, observed on the 30th of August with the Helioseismic and
Magnetic Imager \citep[HMI;][]{HMI} onboard of the Solar Dynamics Observatory
(SDO), which is the same as used in \cite{WB18}. For all the simulation, we use $256\times256\times320$ grid points,
that the horizontal resolutions in the simulation and observation are
the same.
This setup have been successfully used to reproduce the emission
features of coronal active region \citep[e.g.][]{WP19}.

\begin{table*}[t!]\caption{
Summary of runs.
}\vspace{12pt}\centerline{\begin{tabular}{l|crrr}
Run & $H_{\rm M}^{\rm in}$ [G$^2$Mm]&$H_{\rm M}^{\rm bot}$[G$^2$Mm]&$E_{H_{\rm M}^{\rm bot}}$ [G$^2$Mm]&$H_{\rm M\ rms}^{\rm bot}$ [G$^2$Mm]\\[.8mm]   
\hline
\hline\\[-2.5mm]
R          &                        0 &  1.3$\times 10^3$ & 1.5$\times 10^3$  &  2.4$\times 10^3$\\
M0       &                        0 &  2.1$\times 10^3$ & 2.9$\times 10^3$  & 4.7$\times 10^3$\\
M3e3   &  3$\times 10^3$ & 3.7$\times 10^3$ & 1.4$\times 10^3$   & 5.3$\times 10^3$ \\
M1e4   &  1$\times 10^4$ & 1.6$\times 10^4$ & 1.1$\times 10^3$  & 1.6$\times 10^4$ \\
M3e4   &  3$\times 10^4$ & 4.6$\times 10^4$ & 1.6$\times 10^3$  & 4.6$\times 10^4$\\
M1e5   &  1$\times 10^5$ & 1.5$\times 10^5$ & 2.5$\times 10^3$  & 1.6$\times 10^5$\\
M3e5   &  3$\times 10^5$ & 4.7$\times 10^5$ & 8.5$\times 10^3$  & 4.7$\times 10^5$\\
M5e5   &  5$\times 10^5$ & 7.7$\times 10^5$ & 1.6$\times 10^4$  & 7.8$\times 10^5$\\
M1e6   &  1$\times 10^6$ & 1.4$\times 10^6$ & 4.0$\times 10^4$  & 1.4$\times 10^6$\\
\hline
M-3e5 & -3$\times 10^3$& -4.7$\times 10^3$& 9.1 $\times 10^3$ & 4.7 $\times 10^3$\\
\hline
\label{runs}\end{tabular}}\tablefoot{$H_{\rm M}^{\rm in}$ in the
injected magnetic helicity density, the only varying input parameter,
see \Eq{eq:Hin}. $H^{\rm bot}_{\rm M}$ is the horizontal averaged
magnetic helicity at the bottom boundary, the photosphere ($z=0$) with
its error $E_{H^{\rm bot}_{\rm M}}$ determined by the time
variations. $H_{\rm M\ rms}^{\rm
  bot}=<H_{\rm M}^2 (z=0)>_{xy}^{1/2}$ is the horizontal averaged rms
value of the magnetic helicity at the bottom boundary.
All values are averaged over the entire simulation times.
}
\end{table*}

\begin{figure*}[t!]
\begin{center}
\includegraphics[width=2\columnwidth]{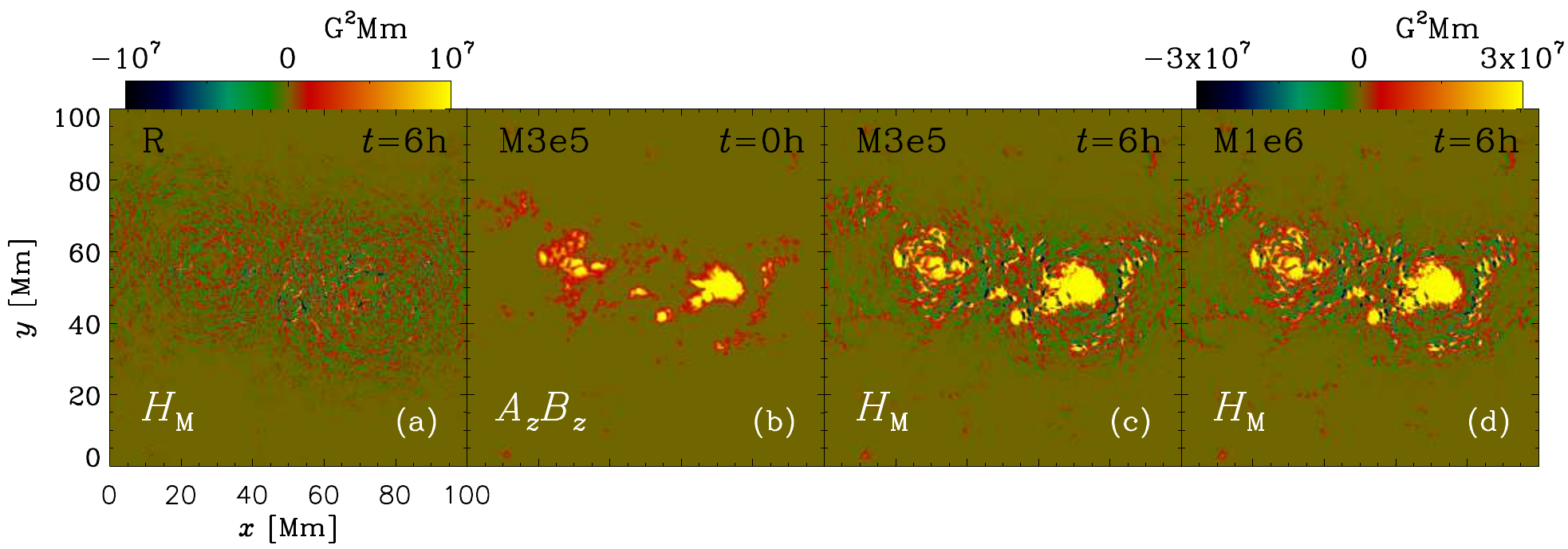}
\end{center}\caption[]{
Magnetic helicity distribution at the photosphere for Runs~R, M3e5 and
M1e6. Panel a shows magnetic helicity $H_{\rm M}$ for Run~R after 6 hours running
time, where no additional helicity is injected. Panel b and c show for
Run~M3e5 the injected helicity at $t=0$ and the helicity after 6 hours running
time, respectively. Panel d shows the same as panel c but for Run~M1e6. 
The color table ranges are the same for panels a-c, see \Sec{sec:hel}.
}\label{maghel_slice}
\end{figure*}

\subsection{Injection of magnetic helicity density}

As we want to study the influence of magnetic helicity on coronae, we
make use of fact the simulations are driven by the photospheric
magnetic field. We can modify this field in such a way that it becomes more
helical without changing the vertical magnetic surface flux. This
allow us to study the effect of magnetic helicity on the corona in an
isolated way.
Throughout this paper, we discussing magnetic helicity, we mean the
magnetic helicity density, defined as
\begin{equation}
H_{\rm M} = \AAA\cdot\BB.
\end{equation}
One advantage of our model consist that we solve the induction
equation in term of $\AAA$ instead of $\BB$, which make $H_{\rm M}$
directly accessible and it can easily be modified.

The observed vertical magnetic field in the photosphere used to drive the simulation is
transformed to $\AAA$ using, see \cite{Bingert2009},
\begin{equation}
\hat{A}_x= i \frac{k_y}{k^2} \hat{B_z} \quad \text{and}\quad \hat{A}_y= -i \frac{k_x}{k^2} \hat{B_z},
\end{equation}
where the hats indicate the Fourier transform of the quantity, $k_x$,
$k_y$ are the horizontal and $k$ the vertical wavenumber with
$k^2=k_x^2+k_y^2$.
$A_z$ is related to horizontal magnetic field and is normally not prescribed
and can evolve freely.
We can now use $A_z$ to include magnetic helicity in photospheric
magnetic field.
For this, we set
\begin{equation}
A_z= {H^{\rm in}_{\rm M}\over<B_z^2>_{xy}} B_z,
\label{eq:Hin}
\end{equation}
where $<\cdot>_{xy}$ indicates horizontal averaging.
$H^{\rm in}_{\rm M}$ is our input parameter for controlling the helicity
injection. If we multiply  with $B_z$ and apply
horizontal averaging on both sides we find that the parameter $H^{\rm in}_{\rm M}$
is directly related to the averaged injected magnetic helicity
\begin{equation}
<A_zB_z>_{xy}= H^{\rm in}_{\rm M}.
\end{equation}
However, the setting of $A_z$ also introduces horizontal magnetic fields,
which then can contribute to magnetic helicity density via $A_xB_x$ and $A_yB_y$.
Therefore, we also define horizontal averaged magnetic helicity at the bottom boundary.
\begin{equation}
 H^{\rm bot}_{\rm M}=<H_{\rm M}(z=0)>_{xy}=<\AAA\cdot\BB>_{xy}(z=0),
\end{equation}
which can be slightly different from the input parameter $H^{\rm
  in}_{\rm M}$ as discussed in \Sec{sec:hel}.
In the following, we use various values of $H^{\rm in}_{\rm M}$ to
drive the corona and investigate their effect on the X-ray production.

\subsection{Magnetic helicity and its gauge}

Before present the results of the numerical simulations, we want to
discus the issues of the gauge dependency of magnetic helicity in our
model. We set the gauge to be the resistive gauge, this means for
$\AAA\rightarrow\AAA+\nab\phi$, we chose $\phi=\eta\nab\cdot\AAA$.
Hence, if we calculate the magnetic helicity in our simulation it is
well defined and consistent among all our simulations. In particular,
as we are mostly interested in the scaling of magnetic helicity to
X-ray emission, the gauge will  not affect our results.
Furthermore, the work of \cite{BSB18} have shown that calculating
magnetic helicity density using the resistive gauge, the magnetic
helicity spectrum and the relative helicity by \cite{BF84} give
consistent results for a simulation of the solar corona, very similar to
the ones used in our work.
We are therefore convinced that calculation of magnetic helicity in
our gauge is meaningful and we can use the observed values of magnetic
helicity \citep{ZBS14,ZBS16} as an motivation to our photospheric
helicity injection.

\begin{figure}[t!]
\begin{center}
\includegraphics[width=\columnwidth]{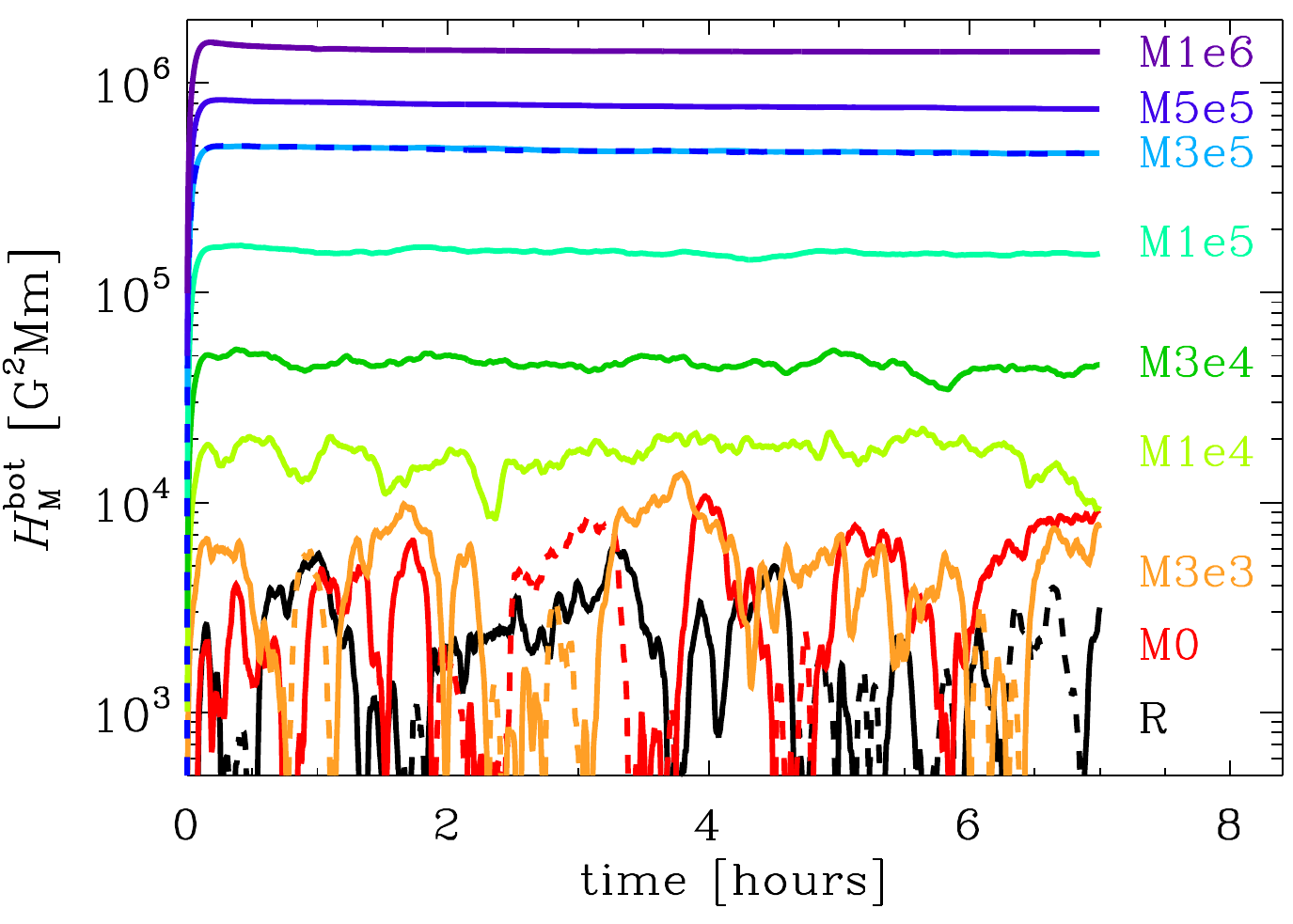}
\end{center}\caption[]{
Horizontal averaged magnetic helicity at the photosphere ($z=0$) $H^{\rm
  bot}$ as a function of time for all runs.
Solid lines show positive values, dashed ones show negative
values. The different runs are distinguished by color as indicate by
their run name on the right side. We note that Run~M-3e5 falls on top
Run~M3e5 with a negative magnetic helicity as indicated by the dashed
line. See \Sec{sec:hel}.
}\label{magh_time}
\end{figure}

\begin{figure}[t!]
\begin{center}
\includegraphics[width=\columnwidth]{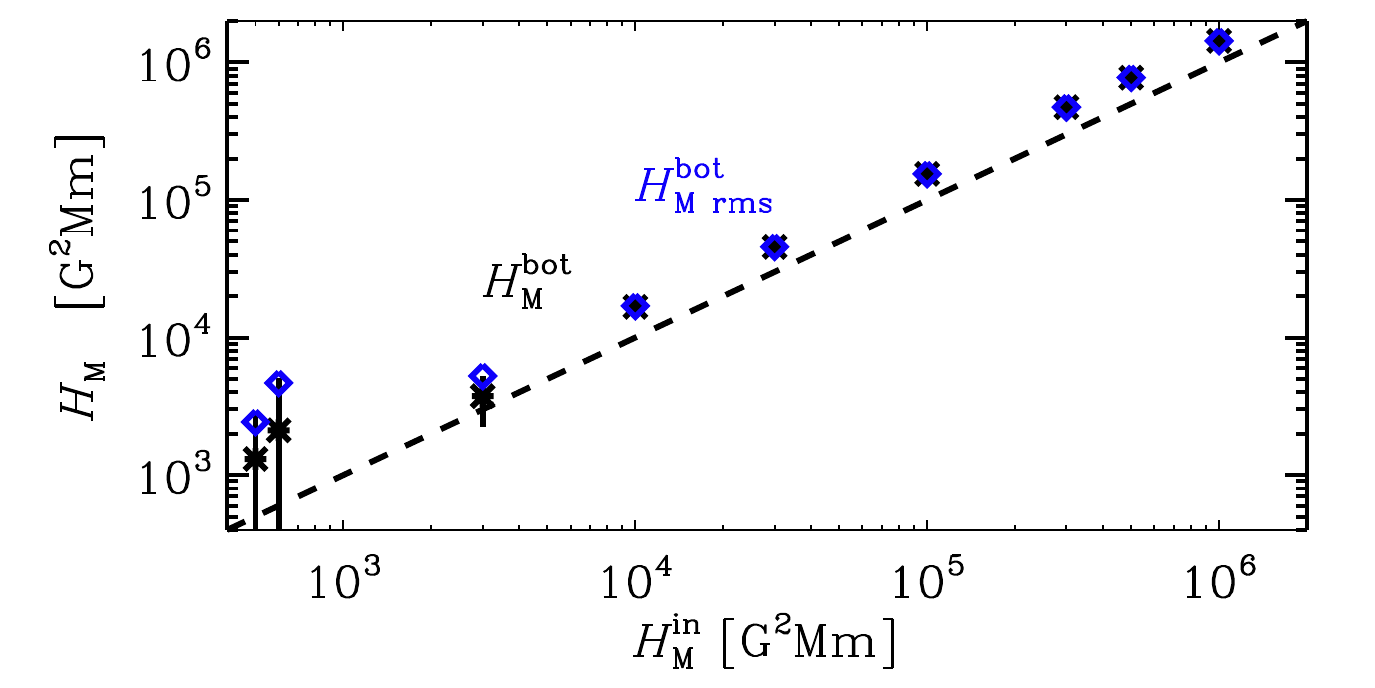}
\end{center}\caption[]{
Magnetic helicity at the photosphere $H_{\rm
  M}$ as a function of injected magnetic helicity $H^{\rm in}_{\rm
  M}$. The black squares show horizontal averaged values and the
blue asterisks show the rms values $H_{\rm M\ rms}^{\rm
  bot}$ at the photosphere ($z=0$), see \Tab{runs} for exact
values. The dashed black line indicate a one-to-one relation.
Runs~R and M0 have been moved to $H^{\rm in}_{\rm M}$=0.05 and 0.06 G$^2$Mm,
respectively, to be able to include these runs in the plot, even
though their values are zero. All values are averaged in time over
entire running time. See \Sec{sec:hel}.
}\label{magh_in}
\end{figure}

\begin{figure}[t!]
\begin{center}
\includegraphics[width=\columnwidth]{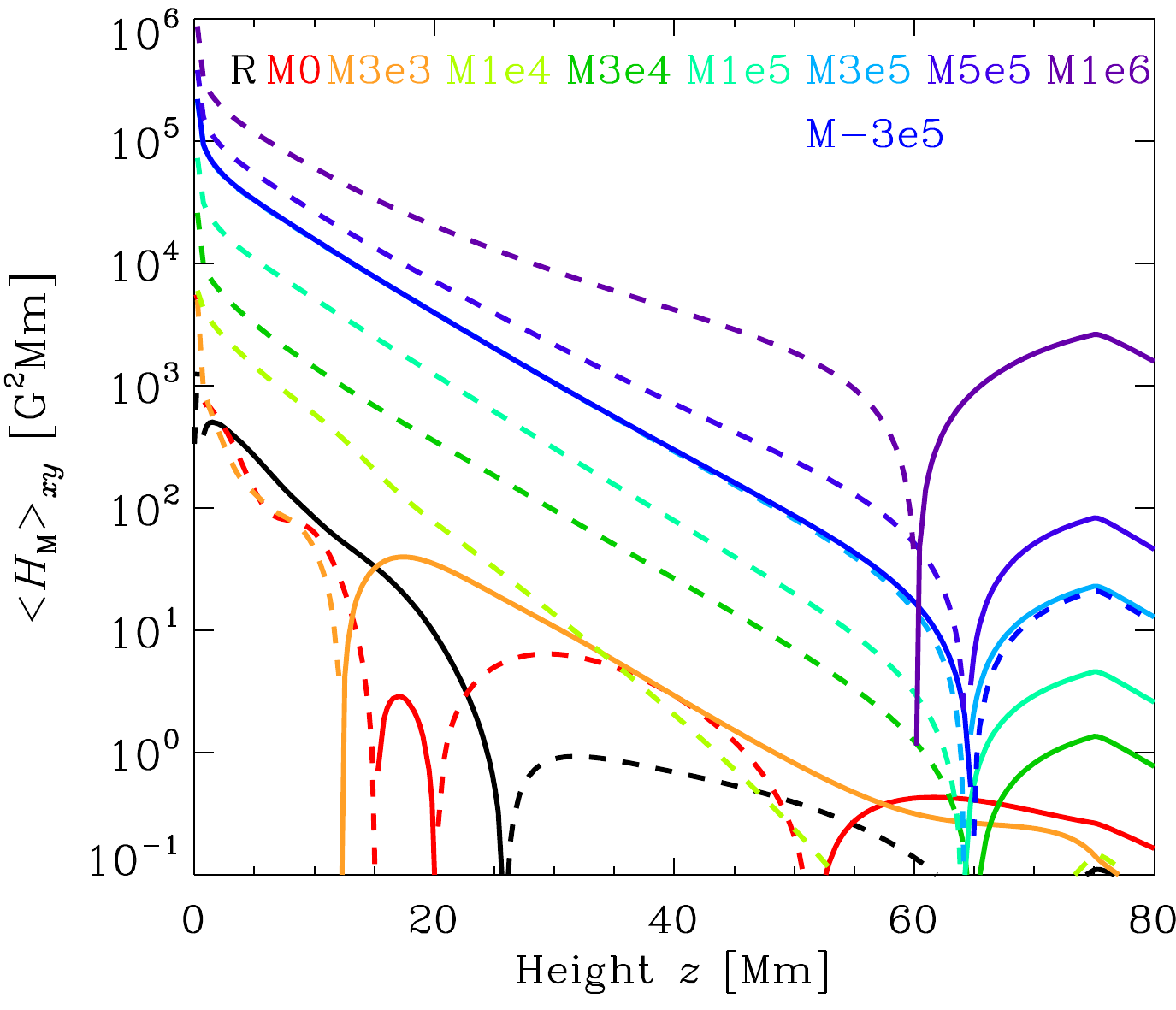}
\end{center}\caption[]{
Horizontal averaged magnetic helicity at a function of height $z$ for
all runs. As in \Fig{magh_time}, solid lines indicate positive values,
dashed indicate negative values. The colors indicate the different
runs. All values are averaged in time over the relax
stage of the runs. See \Sec{sec:hel}.
}\label{maghel_height}
\end{figure}

\section{Results}
\label{sec:results}

For our work we use ten runs where we increase the magnetic helicity injected in
the photosphere. Run~R is the reference run, where $A_z$ in the photosphere is not
modified, hence no helicity injected. This run is similar to Run~Ba of
\cite{WB18}. In all other runs we set $A_z$ according to \Eq{eq:Hin},
The number after 'M' in the run names indicates the value of injected magnetic
helicity in G$^2$Mm, see \Tab{runs}. Run~M0 no magnetic helicity is injected, however,
because $A_z$ is set to zero in the photosphere, the run is different
from Run~R, where $A_z$ can freely evolve.
One run has been injected with a negative magnetic helicity to check
weather or not the sign is important for the amount of X-ray emission.
We chose the values of magnetic helicity motivated by the measurements
of \cite{ZBS14,ZBS16}, where typical active region have values from
10$^{4}$ up to 5$\times10^{5}$ G$^2$Mm.

All the runs have run seven hours to be well in a relaxed stage, in
which the averaged temperature, ohmic heating and X-ray emission do
not change significantly in time. As is it common for such kind of
simulations, the initial phase is dominated by building up a self-heated
corona independent on the initial temperature and density
profiles. This is mostly set by the radiative cooling time in these
models.
For our further analysis we use the time after
4.6 hours as a relaxed stage and calculate most of our results from
this stage.

\begin{figure*}[t!]
\begin{center}
\includegraphics[width=2\columnwidth]{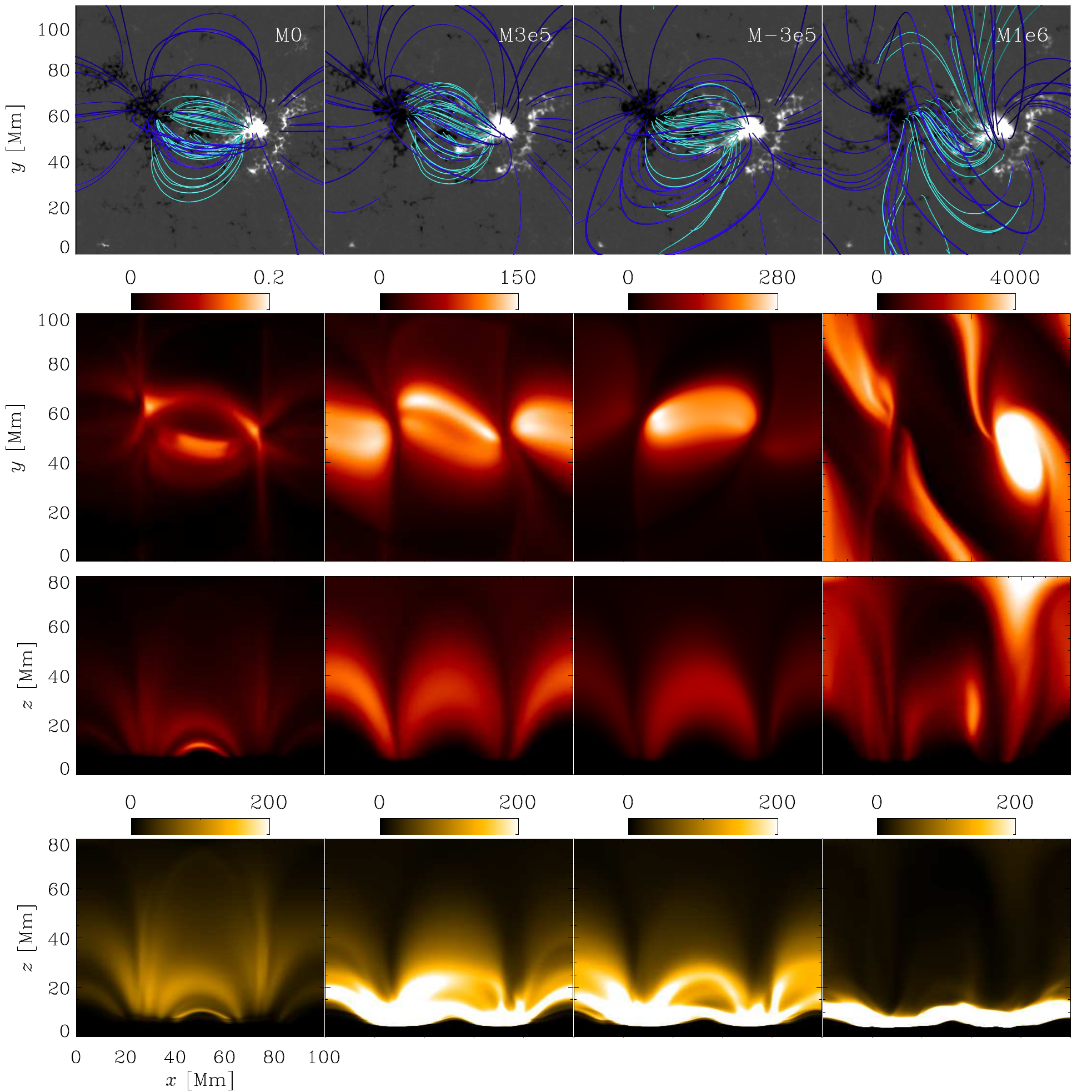}
\end{center}\caption[]{
Magnetic field lines configuration and X-ray and EUV emission for
Runs~M0, M3e5, M-3e5 and M1e6.
The first row shows the vertical magnetic field at the photosphere
(white outward, black inwards, between $-100$ and $100$ G) together
with traced magnetic field lines. The light blue lines show the close
connecting magnetic field between the two polarities and the dark blue
shows the larger arching fields connecting the two. The region
of the seeds for the field line tracing are the same for each runs
The second and third row show the synthesized X-ray emission using the
Hinode/XRT Al-poly temperature response function as top view ($xy$)
and side view ($xz$), respectively.
The last row shows the synthesized EUV emission using the temperature
response function for the AIA~171~\AA\ channel.
The plots have been calculated from a six-hour snapshot of each
simulation.
The emissions are plotted in units of DN pixel$^{-1}$. See \Sec{sec:xray}.
}\label{magf_emis}
\end{figure*}

\subsection{Helicity evolution and distribution}
\label{sec:hel}

First we present the distribution of magnetic helicity injected in the photosphere.
If magnetic helicity is not prescribed via \Eq{eq:Hin}, it is initially
zero. However, because of the photospheric motions interacting with
the photospheric magnetic field, magnetic helicity is generated, as
shown in \Fig{maghel_slice}a for Run~R. 
The helicity is distributed around the two magnetic polarities of the active region
showing both sign of helicity varying around the zero level. The
horizontal averaged value is small and varies over time between
positive and negative values (\Fig{magh_time}).
These values are consistent with calculated ones of \cite{BSB18} for
their simulation of the solar corona.
The time average is small with an error larger
than the value, see third and forth row of \Tab{runs}. The error is
estimated by the largest difference between the mean of each third of
the time series. This means that at each time there a non-zero
magnetic helicity present in the photosphere, however the time average
does not lead to preferred sign.
Run~M0 behaves similarly but with a higher time
averaged value and a larger error.

For the other runs, where we inject helicity, the actual helicity
in the photosphere is the sum of the injected and the by photospheric
motions generated helicity. This can be seen in \Fig{maghel_slice} for
Run~M3e5. In Panel b, we show the injected helicity at beginning of
the simulation $t=0$, when the photospheric motions are zero.
There, the helicity is proportional to $B_z^2$ as given by
\Eq{eq:Hin} and has peak value of $\pm10^7$ G$^2$Mm.
This distribution mimic the actual distribution of helicity in active
region well \citep{ZBS14,ZBS16}.
In the relax stage, see
\Fig{maghel_slice}c,  the by photospheric motions generated helicity
is added to the injected helicity.
As shown in \Fig{maghel_slice}c for runs with even higher magnetic
helicity, the structure does not change, only the amplitudes become
stronger.
Furthermore, the injected helicity via $A_z$ also enhances the
horizontal magnetic field, which lead to additional contribution via
$A_xB_x$ and $A_yB_y$. Hence, the time averaged values are always
around 1.5 times higher than the injected helicity values, see
\Tab{runs} and \Fig{magh_in}.
Only for runs with low magnetic helicity (Runs~R to M3e3) the time
variations are comparable to the mean value, see \Fig{magh_time} and
therefore the rms value is different from the mean value, see
\Fig{magh_in} and \Tab{runs}.
As the mean value and the rms (for most of the runs) is proportional
to the injected value, we can use the injected magnetic helicity
as a reasonable input parameter describing the helicity in the
photosphere well.

Next we look at the height distribution of magnetic helicity in our
simulations.
Interestingly for all of the runs, the magnetic helicity has a different sign in the
corona than in the photosphere, see \Fig{maghel_height}. If we inject
positive helicity, we find negative helicity just above the first grid
layers. This is mostly likely an artifact of how the magnetic field in
terms of the vector potential is set at the boundary. We find this
behavior also in Run~R, so it is not an artifact of the helicity
injection.
Furthermore, the absolute value
of helicity shows a smooth decrease in the grid layers near the
boundary.
The actual sign of helicity do not matter for our study as
it become clear in the analysis below. 

For most of the runs helicity decreases exponentially with
a similar scaling height of around 6-7 Mm. This is similar to the
scaling height of ohmic heating in these kind of simulations
\citep[e.g.][]{GN05a,BP11}.
For runs with low injected helicity (Runs~R to M1e4) the sign of averaged
helicity changes multiple times in the corona showing no distinct pattern.
For larger injected helicities (Runs~M3e5 to M1e6), we find only one
reversal occurring for all runs around 65 Mm. This might be because of
the limited size of the domain.
We do not find any correspondence between location where the sign of magnetic
helicity changes and the location of plasma $\beta=1$ as found by
\cite{BSB18}. 
We think that this is a special property of their simulation and not a
general feature.
However, the fact that the sign changes in the corona of our
simulations agrees well with the observational result that magnetic
helicity has the opposite sign in the heliosphere than on the solar
surface \citep{BSBG11}, which have been also found in helical dynamo
simulations with coronal envelope \citep{WBM11,WBM12} and in solution
of simple dynamos with a force-free corona \citep{Bon16}.

\begin{figure}[t!]
\begin{center}
\includegraphics[width=\columnwidth]{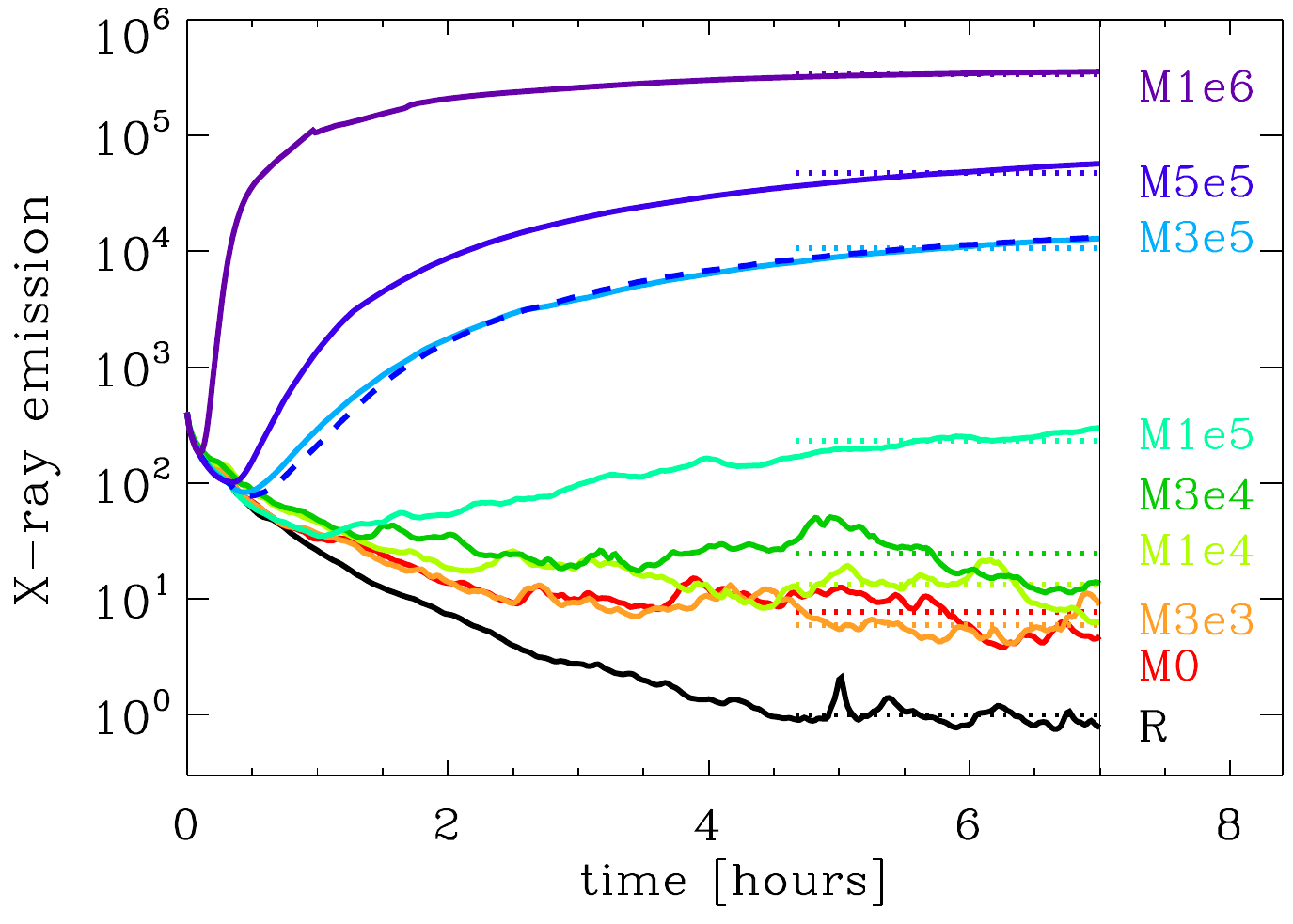}
\end{center}\caption[]{
Time series of the total synthesized X-ray emission for all runs.
The emission is normalized by the one of Run~R in the relax stage.
The dashed purple line shows Run~M-3e5.
The solid vertical line indicate the beginning of the relax stage
(after 4.7 hours). See \Sec{sec:xray}.
}\label{xray_time}
\end{figure}

\subsection{Magnetic field and emission structure}
\label{sec:xray}

\begin{figure*}[t!]
\begin{center}
\includegraphics[width=0.66\columnwidth]{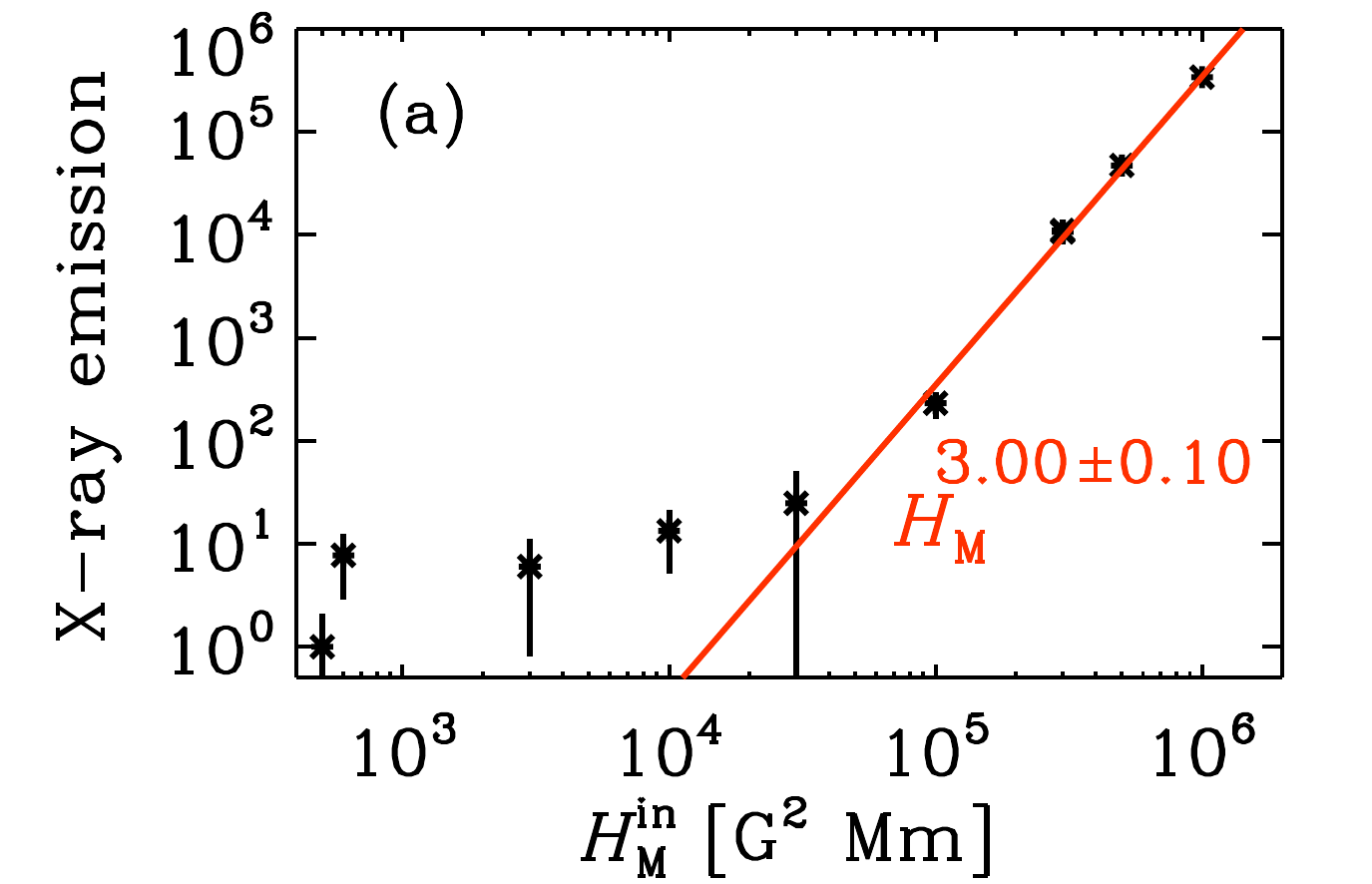}
\includegraphics[width=0.66\columnwidth]{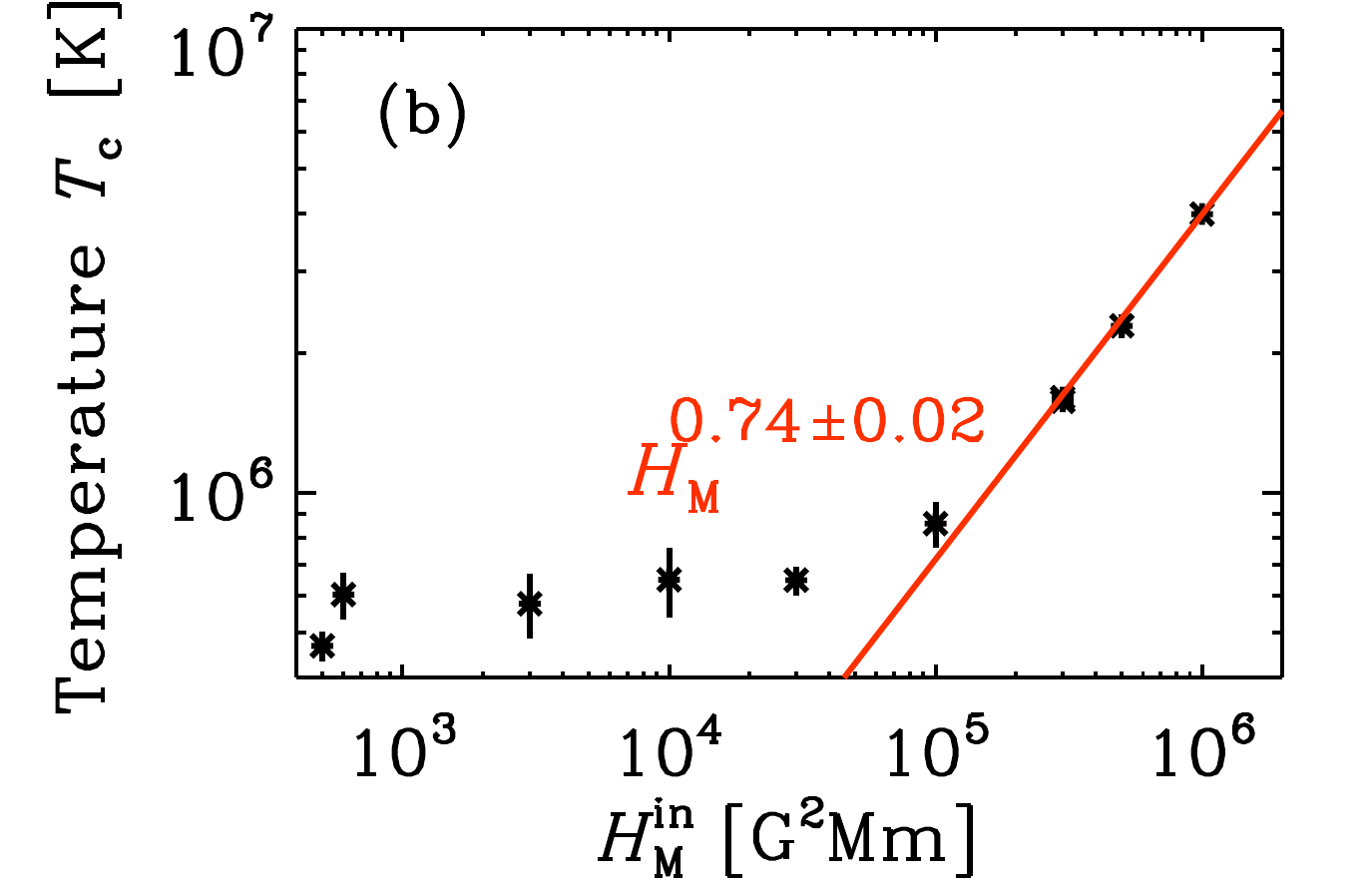}
\includegraphics[width=0.66\columnwidth]{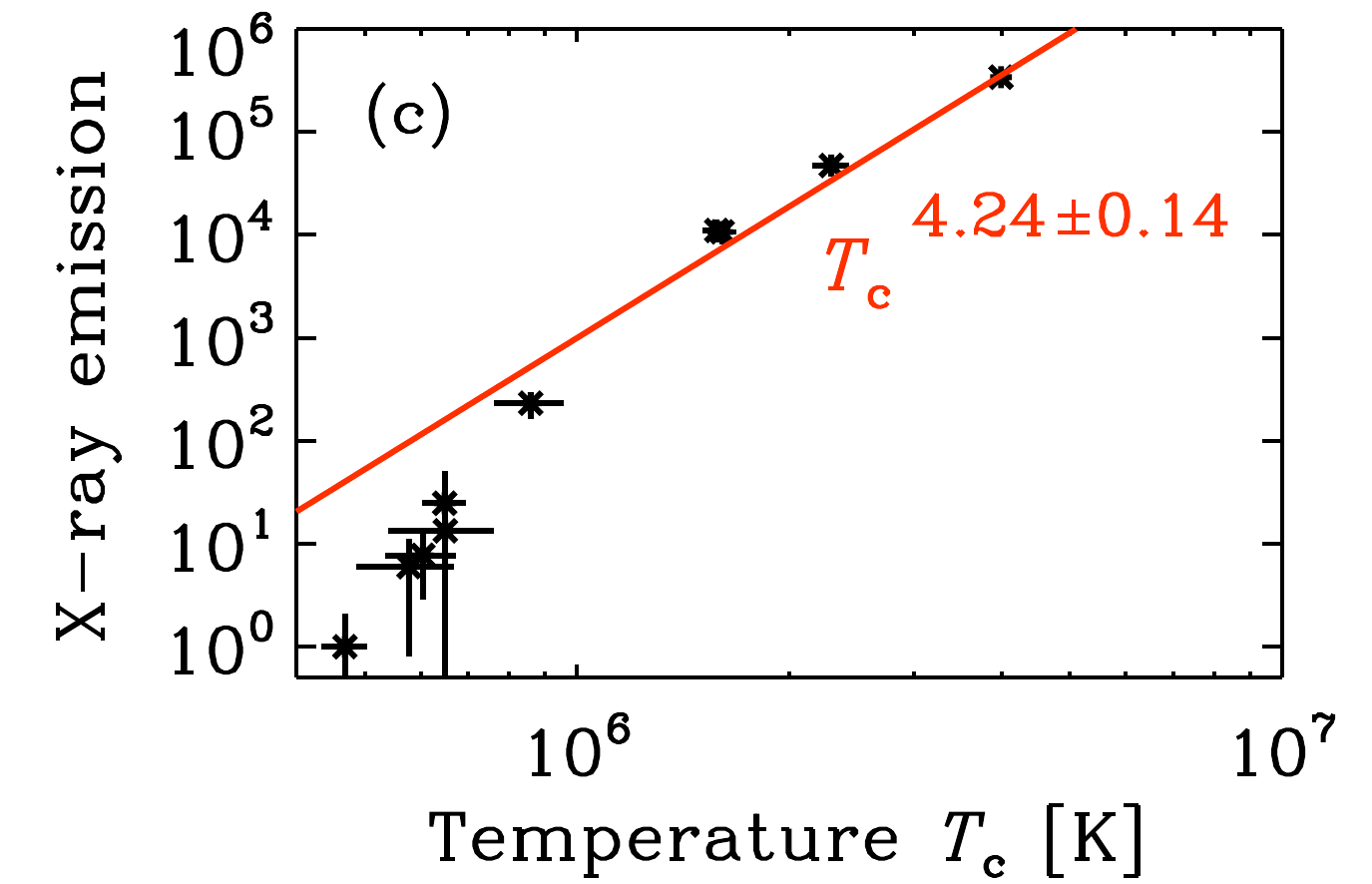}
\end{center}\caption[]{
Scaling of X-ray emission with magnetic helicity and coronal temperature.
We show the total X-ray emission over injected magnetic helicity $H^{\rm in}_{\rm M}$ (a),
coronal temperature $T_{\rm c}$ over $H^{\rm in}_{\rm M}$ (b) and X-ray emission
over coronal temperature $T_{\rm c}$ (c).
The red line is a power-law fit over the last five data points with
corresponding slope in red. The X-ray emission is time-averaged over
the relaxed stage (4.6 to 7 hours) and normalized by total emission of
Run~R. The coronal temperature is averaged in horizontal directions,
in height ($z$=18 to 20 Mm) and over the relaxed stage.
The errors are estimate from the time variation in the relaxed stage.
As in \Fig{magh_in} we moved Runs~R and M0 to $H^{\rm in}_{\rm M}$=0.05 and 0.06 G$^2$Mm,
respectively, to include them in the plot. See \Sec{sec:scale}.
}\label{xray_scale}
\end{figure*}

The injection of magnetic helicity at the photosphere affects the
magnetic field structure in the corona of each simulations. 
As shown in the top row of \Fig{magf_emis}, the vertical magnetic
field in the photosphere do not change, if magnetic helicity is
injected. However, field line topology visible as a field line twist undergoes significant
changes due to various levels of magnetic helicity in the photosphere.
For Run~M0 (and similar for Run~R), the magnetic field lines show
potential-like arch structures. 
Increasing magnetic helicity let the field lines be
become more helical forming sigmoid-like shapes.
For the larges helicity input (Run~M1e6), the field lines are forming
large complex arcs instead on small close connecting arcs, which are
most pronounce in the light blue field lines, see first row of
\Fig{magf_emis}.
We note here that due to the periodic boundary condition the
field lines can connect through the horizontal boundary.
As expected, an opposite sign of helicity forms a sigmoid with the
opposite handiness. Except the handiness, the arc structures does not
depend on the sign of magnetic helicity, as the magnetic
field line topology of Run~M-3e5  is an mirror image of the ones of Run~M3e5. 
The small differences can be mostly associated to the field line tracing
algorithm and the non-symmetry of the active regions.

The main goal of this work is to the relate the X-ray emission to the
injected magnetic helicity. For this we synthesis the X-ray emission
as it would be observed with the X-ray Telescope \citep[XRT:][]{XRT} on the HINODE
spacecraft \citep{Hinode} using the Al-poly channel. We use the density and
temperature of the model to calculate the emission using the
temperature response function calculated with the help of CHIANTI
database \citep{Chianti1,Chianti8}.
This emission calculation assumes an optical thin solar corona and is
either integrated in the $z$ or $y$ direction to mimic a line-of-sight
and a side view, respectively. 
We find magnetic helicity affects strongly the X-ray emission.
As shown in the second and third row of \Fig{magf_emis} for the top ($xy$)
and the side ($xz$) view of the synthesis X-ray emission larger
helicity input lead to higher X-ray emission.
For no magnetic helicity input the X-ray emission is very weak and
a result of a low temperature corona with a potential-like field structure.
For higher magnetic helicity the
X-ray emission becomes significantly stronger and also the loop
structure changes.
For Run~M0 the weak emission comes from small low connecting loop, but is so
weak that the XRT would never able to detect it.
For Runs~M3e5 and M3e-5, the
X-ray emission show twisted sigmoids-like loop structures, aligned with the
magnetic field topology seen in the row above. For the high magnetic helicity case of
Run~M1e6 the X-ray emission reveals highly twisted loops structures,
which are significantly deformed by the twist of the magnetic field.
The twist is even so strong that it causes the magnetic field to interact
with the upper boundary and forms a locally enhanced ohmic heating
region, which produce strong X-ray emission.
Even though this emission region is an artifact, its effect on the
averaged total emission is comparable to the temporal variation inside
the relaxed stage and therefore do not affect our results of X-ray
emission scaling below.
Already from these plot we find that the  X-ray emission increases
larger than linear for increasing magnetic helicity.

The extreme UV emission is also affected by the injection of magnetic
helicity. We focus on the bands in extreme UV emission as they would
be observed by the Atmospheric Imaging Assembly
\cite[AIA;][]{LTAB2012}.
Similar as for  the X-ray emission calculation we use the temperature and
density in each simulation together with temperature response kernel
\cite[][]{AIA:2012}\footnote{Implemented in SolarSoft
  (\url{http://www.lmsal.com/solarsoft/}).} of several AIA channels.
As an example we show in the last row of \Fig{magf_emis} the emission
of the AIA~171~\AA\ channel.
Runs with higher magnetic helicity show not only more emission,
but also with a different structure. 
For Runs~M3e5 and M3e-5, the emission comes from coronal loops
structures which also radiates in the X-ray band, however the EUV
seems to be slightly at lower heights than the X-ray emission.
In contrast, for Run~M1e6, the EUV complete vanish from the coronal loops
structures and emits only from the coronal base at around $z$=10 Mm.
We discuss this issue in more detail, when we compare the scaling of
total emission of EUV and X-ray emission with helicity, see
\Sec{sec:euv}.

As our analysis is focused on the X-ray emission, we use this emission to
determine the relaxed stage of our simulations.
For this we plot in \Fig{xray_time} the time evolution of the total
X-ray flux, integrated over the entire simulation domain.
For all the runs the total X-ray emission becomes quasi-steady in time
after around 4.6 hours. We define the quasi-steady stage as a stage
when the temporal variation are small compare to the average. We call
this time interval the relaxed stage and apply all the time averages
over it, which are use in the analysis below.
The duration of the initial phase is determine by the cooling
time in system. As the Run~R has a very low temperature corona the
cooling times is therefore much longer than for the other runs, as
seen also in the X-ray emission.
Already in this plots we also see an indication of a power-law
relation with an index larger than one between the X-ray emission and the injected magnetic helicity.

\begin{figure*}[t!]
\begin{center}
\includegraphics[width=0.66\columnwidth]{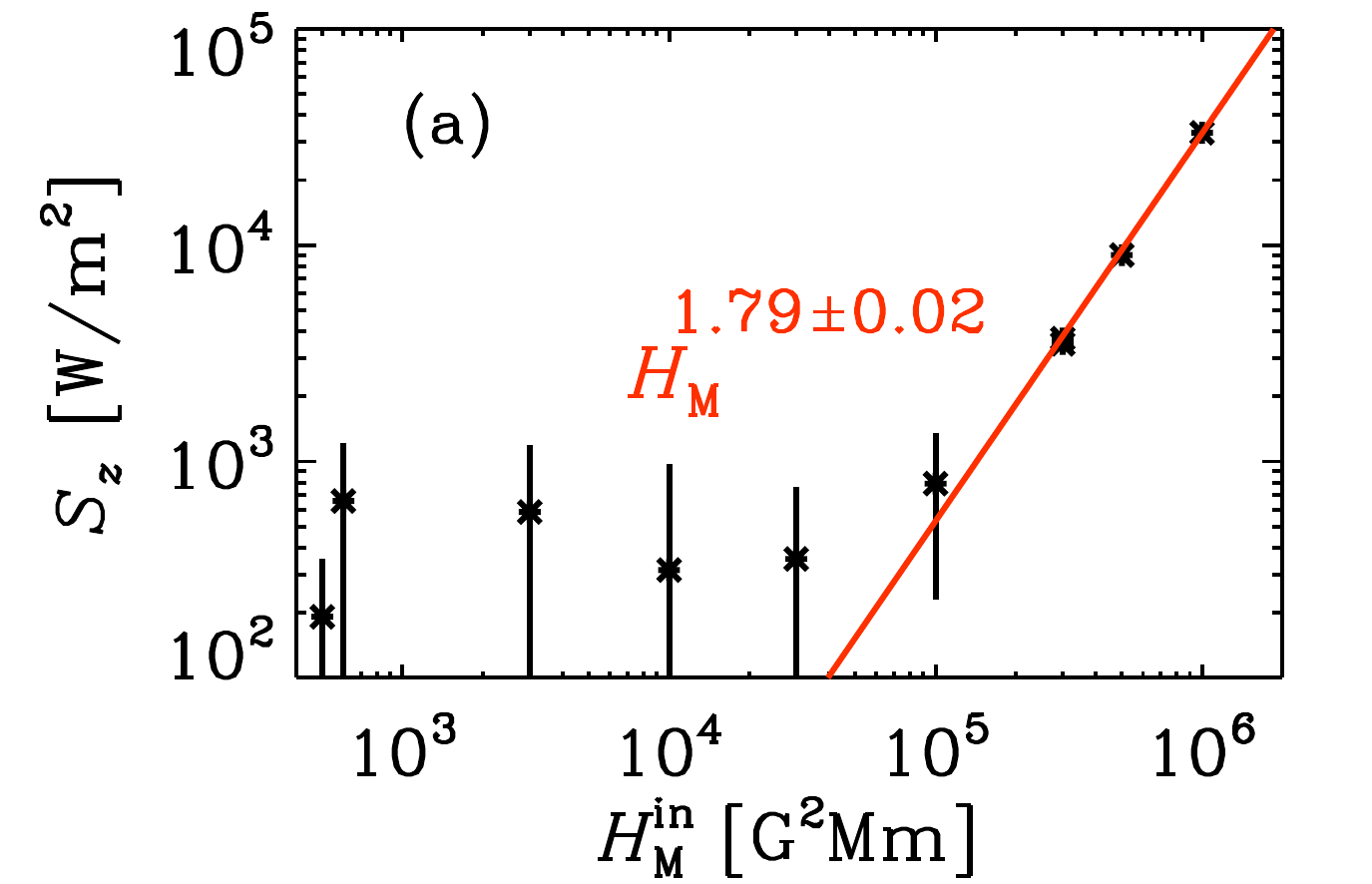}
\includegraphics[width=0.66\columnwidth]{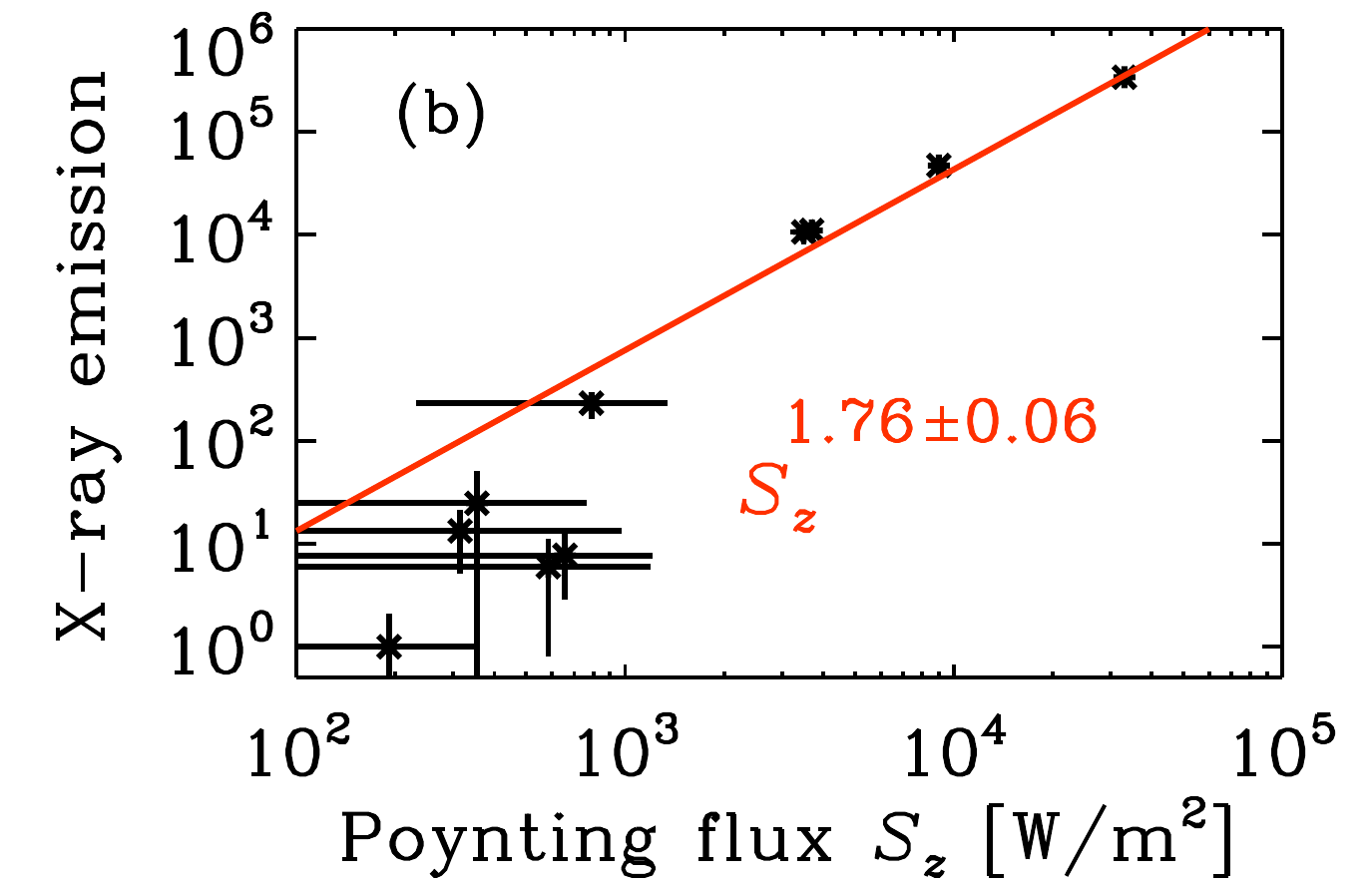}
\includegraphics[width=0.66\columnwidth]{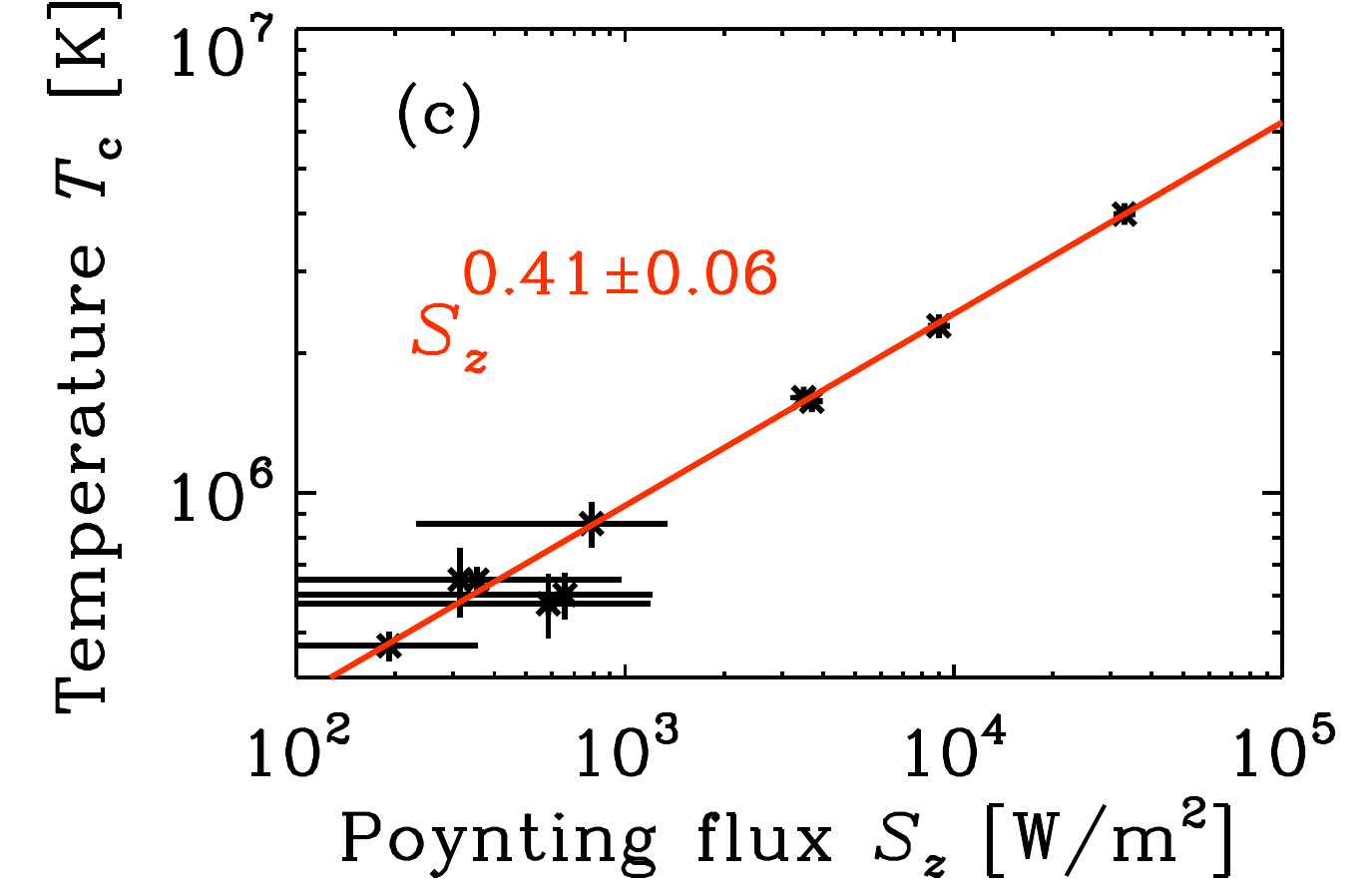}
\end{center}\caption[]{
Scaling of Poynting flux with magnetic helicity, X-ray emission and coronal temperature.
We show the vertical Poynting flux $S_z$ over injected magnetic helicity $H^{\rm in}_{\rm M}$ (a),
X-ray emission over vertical Poynting flux $S_z$ (b) and coronal temperature $T_{\rm c}$ over vertical Poynting flux $S_z$ (c).
The red line is a power-law fit over the last five data points with a corresponding slope in red. The vertical Poynting flux is calculated
in the region, where plasma $\beta$ is close to unity ($z=3-4$ Mm) and averaged horizontally and over the
relaxed stage. Otherwise the same as \Fig{xray_scale}.  See \Sec{sec:scale}.
}\label{poyn_scale}
\end{figure*}

\begin{figure}[t!]
\begin{center}
\includegraphics[width=\columnwidth]{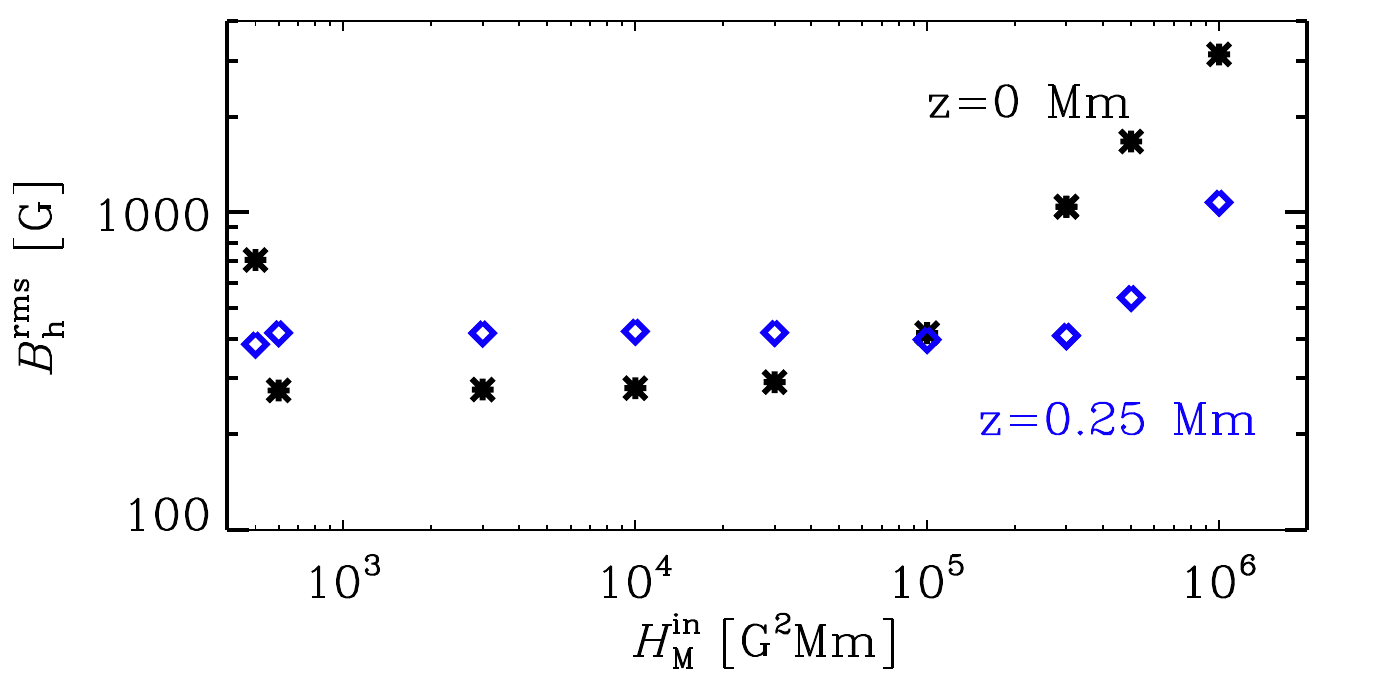}
\end{center}\caption[]{
Horizontal magnetic field $B_{\rm h}^{\rm rms}$ near the photosphere and injected magnetic
helicity $H^{\rm in}_{\rm M}$.
We plot rms values of the horizontal magnetic field at $z=0$ (black asterisks) and
$z=0.25$ Mm (blue diamonds) over $H^{\rm in}_{\rm M}$.
We moved Runs~R and M0 to $H^{\rm in}_{\rm M}$=0.05 and 0.06 G$^2$Mm,
respectively, to include them in the plot.  See \Sec{sec:scale}.
}\label{maghel_brms}
\end{figure}

\subsection{Scaling of X-ray emission}
\label{sec:scale}

Now we turn to question how total integrated X-ray emission depends on
the injected magnetic helicity.
As shown in in \Fig{xray_scale}a, we find a clear increase of X-ray emission for larger injected
helicity. For runs with helicities starting with 3e5 G$^2$Mm we find a
power-law relation with a slope of 3.  For lower helicities the increase
is much lower. This indicates that the helicity has to overcome a
threshold of around 3e5 G$^2$Mm before influencing the X-ray emission significantly.
This relation is tightly
connected to the increase of coronal temperature $T_{\rm c}$ with injected helicity as
shown in \Fig{xray_scale}b. There we find that the coronal temperature
increase with helicity with a power-law exponent of 0.74. for the runs
with higher helicity. The temperature is averaged in horizontal directions,
in height ($z$=18 to 20 Mm) and over the relaxed stage to describe a
typical coronal temperature of each simulation.
Also here a threshold of  around 3e5 G$^2$Mm is present.
The power law relation of X-ray emission and temperature of $T_{rm c}^{4.24\pm0.14}$
follows closely the relation obtained from stars $T^{4.5\pm 0.3}$
\citep{guedel:2004} and is close the relation of $T^4$ expected from
thermal black body radiation.
However, we should keep in mind that also the density enters the
calculation of X-ray emission. Hence, the part of the increase of
X-ray emission is also due to the increase of density inside the
emitting X-ray loops.
Getting a consistent temperature X-ray relation make us confident we
can use our results to understand the
stellar X-ray emission as a function of helical magnetic fields.

Before discussion the implication of this result for stellar
activity we investigate why magnetic helicity actually
increases the X-ray emission in the way as we found in our
simulations.
First of all, as discussed above the X-ray can be directly
related to the coronal temperature either by using the observed
relation or even a simple black boundary radiator, where the emission
is proportional to the temperature to forth power.
Our runs with high magnetic helicity are well in agreement with these two relations.
The coronal temperatures in all simulations can be
directly related to vertical Poynting flux into the corona, as the
energy flux going into the corona must the be same as the energy flux
going out of the corona. The higher the Poynting flux, the higher the
coronal temperature.
The Poynting flux in our simulation is given
\begin{equation}
\SSS=\eta \JJ\times\BB-{1\over\mu_0}\left(\uu\times\BB\right)\times\BB,
\label{eq:poynt}
\end{equation}
where the first term in only important in the few lowest grid layers.
In \Fig{poyn_scale}a we show the vertical Poynting flux $S_z$ at the height, where
plasma $\beta$ is around unity, over magnetic helicity.
Plasma $\beta$ is defined as the ration of gas and magnetic
pressure and normally indicated the dominance of magnetic (thermal) energy as low (high) values.
We chose this height, because there the flow field can still dominate the
magnetic field evolution and the Poynting flux give a good indication
for the energy flux into the corona.
The vertical Poynting flux increases with injected helicity for last
four runs with power-law relation of around 1.8.
For runs with low helicity the Poynting flux is not or only weakly increasing helicity.
This results in a relation of X-ray emission to Poynting flux with a
slope of around 1.8 for runs with high helicity and an inconclusive
relation for runs with low helicity, see \Fig{poyn_scale}b.
However, the coronal temperature is can be related to the Poynting flux with a
power-law including all runs, meaning this relation is general and not
dependent on helicity.
Hence, the reason why the runs with lower helicity do not show a
strong increase of coronal temperature with helicity must be due to the 
relation of helicity to Poynting flux.
The Poynting flux in our simulation consists mostly of the horizontal
motions interacting with the horizontal and vertical magnetic field.
The vertical magnetic field and the horizontal motions are the same
for all simulations, because they are prescribed at the photospheric boundary.
Therefore, the magnetic helicity changes the Poynting flux via the
horizontal magnetic field.
As plotted in \Fig{maghel_brms}, the root mean squared of the
horizontal magnetic field $B_{\rm h}^{\rm rms}$ near the photosphere
changes only above the threshold of 3e5 G$^2$Mm.
The discrepancy between the height, where the Poynting flux and the
horizontal magnetic field is calculated are not significant.
The Poynting flux near the photosphere follow a similar behavior as at
$z=3-4$ Mm, where plasma $\beta$ is unity; there, the values of
the time-averaged Poynting flux are close to zero or even negative,
therefore the scaling relation as shown in \Fig{poyn_scale}a is not
possible to determine.
Runs~R and M0 have also different values of horizontal magnetic
fields, because $A_z$ in the photosphere is set to zero in Run~M0 but
not set in Run~R, see
\Fig{poyn_scale}a and \Sec{sec:hel}. This might also explain their
different X-ray emission and coronal temperatures for similar values
of photospheric helicity.
To conclude, adding magnetic helicity to a photospheric magnetic field
has only a significant effect, if the resulting horizontal magnetic
field exceeds the exiting horizontal field. 

The fact that X-ray emission is sensitive to the amount of magnetic
helicity at the photosphere can have a large impact on stellar
activity.
Magnetic helicity related to active region is thought to be produced
by the underlying dynamo. Using mean-field theory, the $\alpha$ effect
produces both sign of magnetic helicity, one sign of large scales, the
other on small scales. \citep{BS05}. The large scales in this context
means the scale of the whole stars, small scales everything
significantly below. Hence, magnetic field from active region is
counted as a small scale field in this view. However, as the sign for
the results of our work is not important, for our estimate it does not
matter, if the magnetic helicity injected in our simulation is
produced at large or small scales.
The amount of helicity produced on both scales is $2\alpha\meanBB$ \citep{KR80},
where $\meanBB$ is the large scale field. The $\alpha$ is proportional
to the rotational influence of the Star as it its is proportional to
the kinetic helicity \citep{SKR66}.
However, recent measurement from global convective dynamo
simulations indicate that $\alpha$ has a more complicated distribution
and a different values than the theoretical expression invoking the
kinetic helicity \citep{WRTKKB17,VKWKR19}. Furthermore, recent studies found
a scaling of the $\alpha$ effect with rotational influence to the
power of around 0.7 \citep{W18} or 0.5 \citep{WKK19}.
If we now use this scaling of magnetic helicity with rotational
influence together with our scaling of X-ray emission with magnetic
helicity, we get that the X-ray emission scales with the power of
around 3 using the theoretical scaling of $\alpha$ or around 2 to 1.5
using the $\alpha$ effect measurements.
All values predict are large impact of the magnetic helicity in the
relation of X-ray emission and rotation of stars. Increasing the
magnetic helicity alone without the magnetic flux is enough to
reproduce a scaling similar or even larger what is found be stellar
observation studies \citep[e.g.][]{PMMSBV03,VGJDPMFBC14,RSP14,WD16},
in which the scaling is around 2.

\subsection{Relation to extreme UV emission}
\label{sec:euv}

\begin{figure}[t!]
\begin{center}
\includegraphics[width=\columnwidth]{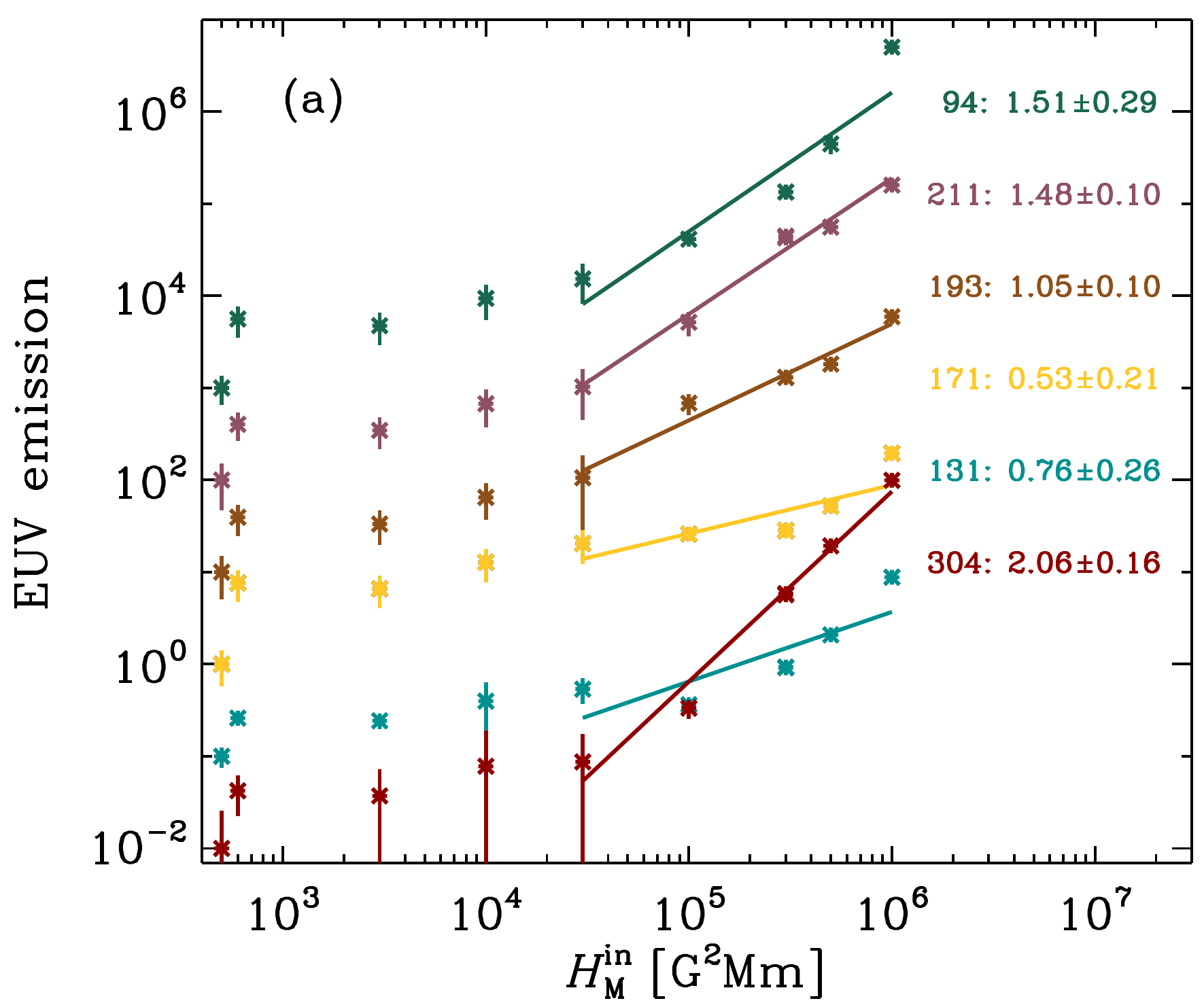}
\includegraphics[width=\columnwidth]{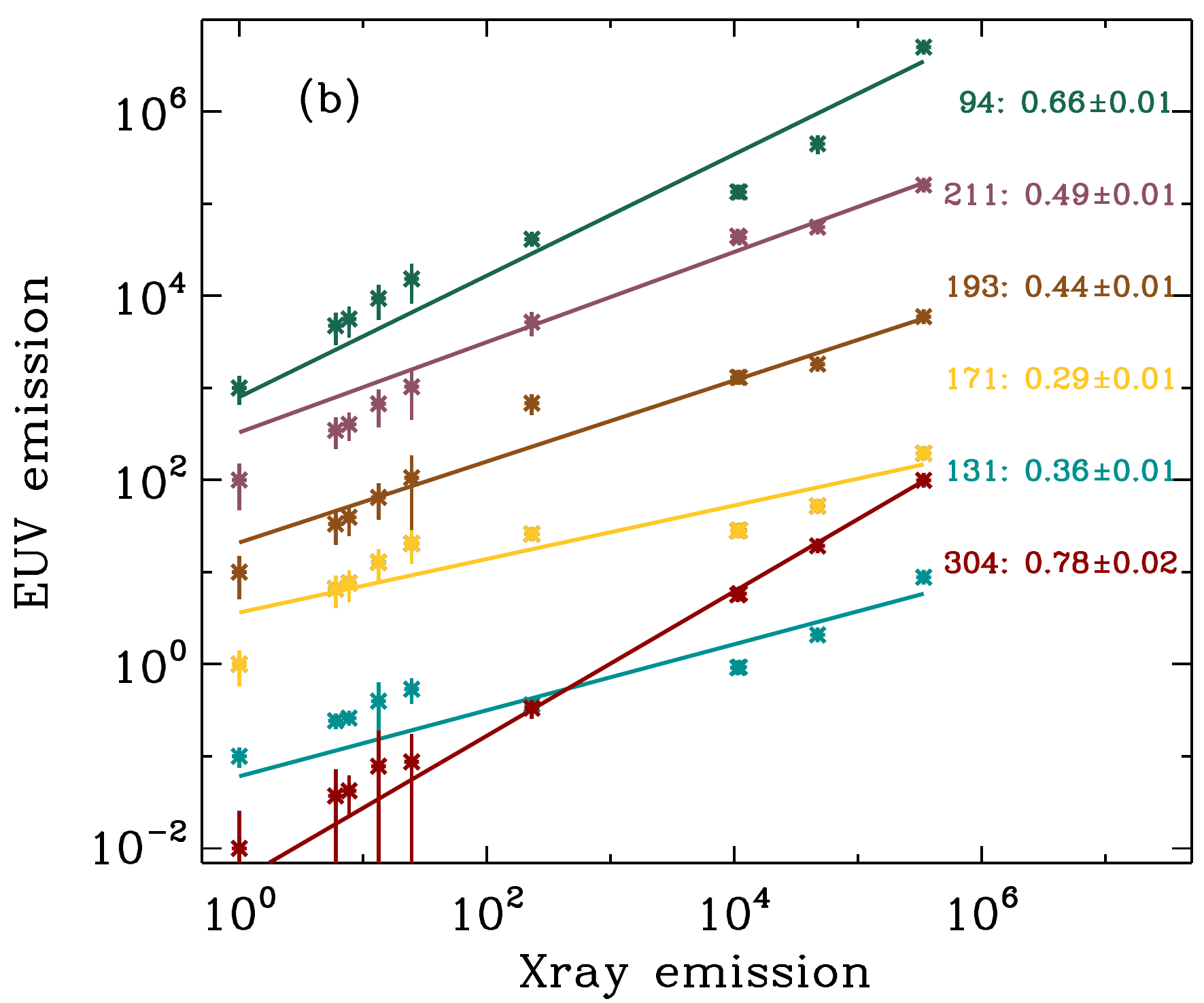}
\end{center}\caption[]{
EUV emission as a function of injected magnetic helicity $H^{\rm
  in}_{\rm M}$ (a) and X-ray emission (b).
Total EUV emissions of the AIA channels 94, 211, 193, 171,
131, 304 \AA\ are indicated by their corresponding wavelength in \AA\
and color. We fit a power-law fit to the last five points (a) or all
points (b) and give the slopes next to their channel number.
All points have been normalized to the emission of Run~R and then
shifted to get a factor of ten ration between the set of points for
each emission to increase visibility. We have ordered the emission based on their peak
temperature with the highest on the top.
We moved Runs~R and M0 to $H^{\rm in}_{\rm M}$=0.05 and 0.06 G$^2$Mm,
respectively, to include them in panel a.
}\label{EUV}
\end{figure}

Beside the X-ray emission we investigate how the extreme
UV emission depends on the magnetic helicity.
EUV emission cannot be observed from other stars than the Sun, however we
can use these emission measures to probe the atmospheric properties
for different kind of stellar activity levels.
For this we synthesized the main AIA coronal channels as described in
\Sec{sec:xray} and plot them as function of injected magnetic
helicity in \Fig{EUV}a.
All EUV emissions show an increase with higher injected helicity. Their
slopes determined from the last five highest helicities are weaker
than the increase of X-rays as shown in \Fig{xray_scale}a.
The slopes show a clear temperature dependence. For high temperatures
emission as in the 94 and 211~\AA\ channels the increase goes with the power of around
1.5, this slope decreases for temperatures around 1 MK (171~\AA\ channel) and then increases again for temperatures of around
200.000 K (304~\AA\ channel).
This show each channel has a significant different magnetic helicity
sensitivity.
As seen in the scaling of X-ray emission, temperature and Poynting with
magnetic helicity, only the five highest magnetic helicity cases show
a clear power-law behavior.
This behavior can be partly explained by the different temperature
response function of each emission channel.
Additionally, the temperature does also effect the density
distribution in each simulation making coronal loop denser for higher
temperatures and less dense for lower temperature. This will than also
effect the synthesis emission calculations.
From measurements of emission from the Sun and other stars, it is found
that the emission from higher temperatures are more sensitive to the
surface magnetic field. \citep[e.g.][]{SCZS89,KPCS18}. This seems to be also true for
the sensitivity of magnetic helicity. However, the difference is that
the 304~\AA\ channel is more sensitive than the slightly higher
temperature.
This could be an artifact, as at the temperatures and heights, where
this emission is radiated, the atmosphere is not fully optical thin any
more.

Of further interest is the relation between different emission
channels, the is often called a flux-flux relation referring to the
relation of different emission fluxes \citep[e.g.][]{Sch87}.
In \Fig{EUV}b we relate the X-ray emission to several EUV channels.
Interestingly, we find power-law relations, which fits all runs
and not just to the ones with highest magnetic helicity. This again
means, that the general emission properties are similar in each of the
simulation and just the relation between magnetic helicity and
Poynting flux. One can also say that the horizontal magnetic field is the
quantity, which causes the difference between the low helicity runs and
the high helicity runs.
Also in the relation between X-ray emission and EUV emission we find
that the slope depends strongly with temperature. At the high and low
temperature ends, we have larger slopes and weaker ones for around
1MK.
As a consequence, the relation of \Fig{EUV} indicate that the magnetic
helicity is more sensitive to X-ray emission rather than EUV emission.
Hence, magnetic helicity related phenomena can be easier
traced with X-ray emission than with other EUV emission channels.
To compare one of the channels with stellar observation we can use the
304~\AA channel, which probe emission from the chromosphere and
transition as a proxy for the C{\ion IV} emission, routinely measure for other stars.
\cite{Sch87} find that the X-ray flux depend on C{\ion IV} with a power
of 1.5, which is reasonable close to $1/0.78=1.28$ of our findings.
However, more recent measurement of chromospheric Ca II H\&K emission
and its dependence on X-ray flux gives a power law of 0.4 \citep{MSS18},
which is closer to our scaling  of the emission in the 131 and
171~\AA\ channel than the 304~\AA\ channel.

Another aspect in the comparison of X-ray emission and EUV emission is
the their location of emission. As shown in \Fig{magf_emis}, the X-ray
emission is emitted from the coronal loop structures for medium and
large helicities, the EUV on the other hand is emitted from the
coronal loops only at the medium helicities. Interestingly if one
compares the decrease of emission with height for these different runs,
see \Fig{emis_height}, one finds that the emission falls off with a
similar scale height for moderated helicities. However, for large
helicities the emission is strongly concentrated in the low corona.
This is in contradiction to the emission signatures found in solar
active regions with high activity. Even though there also the X-ray
emission reaches high values and form loops structures similar in our
simulations, the EUV emission is still visible in large part of the
corona and forming also loops.
This point into the direction that the Sun do not form very active
regions in the same process as discussed here, however, the active
region driving our simulations in the photosphere is also a very
simple one, in contrast to strong active regions on the Sun, which are
often very complex.

\section{Conclusions}
\label{sec:con}

For the first time, we have used numerical 3D MHD simulations of the solar
corona to study the effect of magnetic helicity on the coronal
properties in particular the coronal X-ray emission.
We found that for higher injected magnetic helicity at the photosphere
the corona becomes hotter and emits significantly more X-ray
emission. We determine that the X-ray emission increases with cubicly
with the injected helicity.
Using this scaling with the typical scaling of magnetic helicity on
rotation, we find that can explain the increase of stellar X-ray
luminosity with rotation alone by the increase of magnetic helicity
with rotation.
Therefore with this work we can show that the magnetic helicity can
have a larger impact on the rotation-activity relation of stars.

The increase of magnetic helicity increases the horizontal magnetic
field in the photosphere and therefore the photospheric Poynting flux.
The higher Poynting flux leads to higher coronal temperatures and
larger X-ray flux, followings a $T_{rm c}^{4.24\pm0.14}$ relation in
agreement with stellar observations \citep{guedel:2004}.
Furthermore, we found that the increase of magnetic helicity reproduce
the scaling of chromospheric \& transition zone emission well as stars
predicted.

We find that the reversal of magnetic helicity in the corona is not
related to the location, where plasma $\beta$ is unity, in contrast
to the work of \cite{BSB18}. However, horizontal averaged helicities show at least
one reversal in the corona of each of our simulation and is therefore
in agreement with the studies by \cite{BSBG11} and \cite{WBM11,WBM12}.

Future studies will include the injection of helicity with more
complex distribution of values and signs. Furthermore, we will test
the different between the injection of helicity with our method and
the injection of helicity flux as used observationally studies
\citep[e.g.][]{CWQGSY01,NZZ03,MKYS05,V19}.

\appendix
\section{Emission over height}
\begin{figure}[t!]
\begin{center}
\includegraphics[width=\columnwidth]{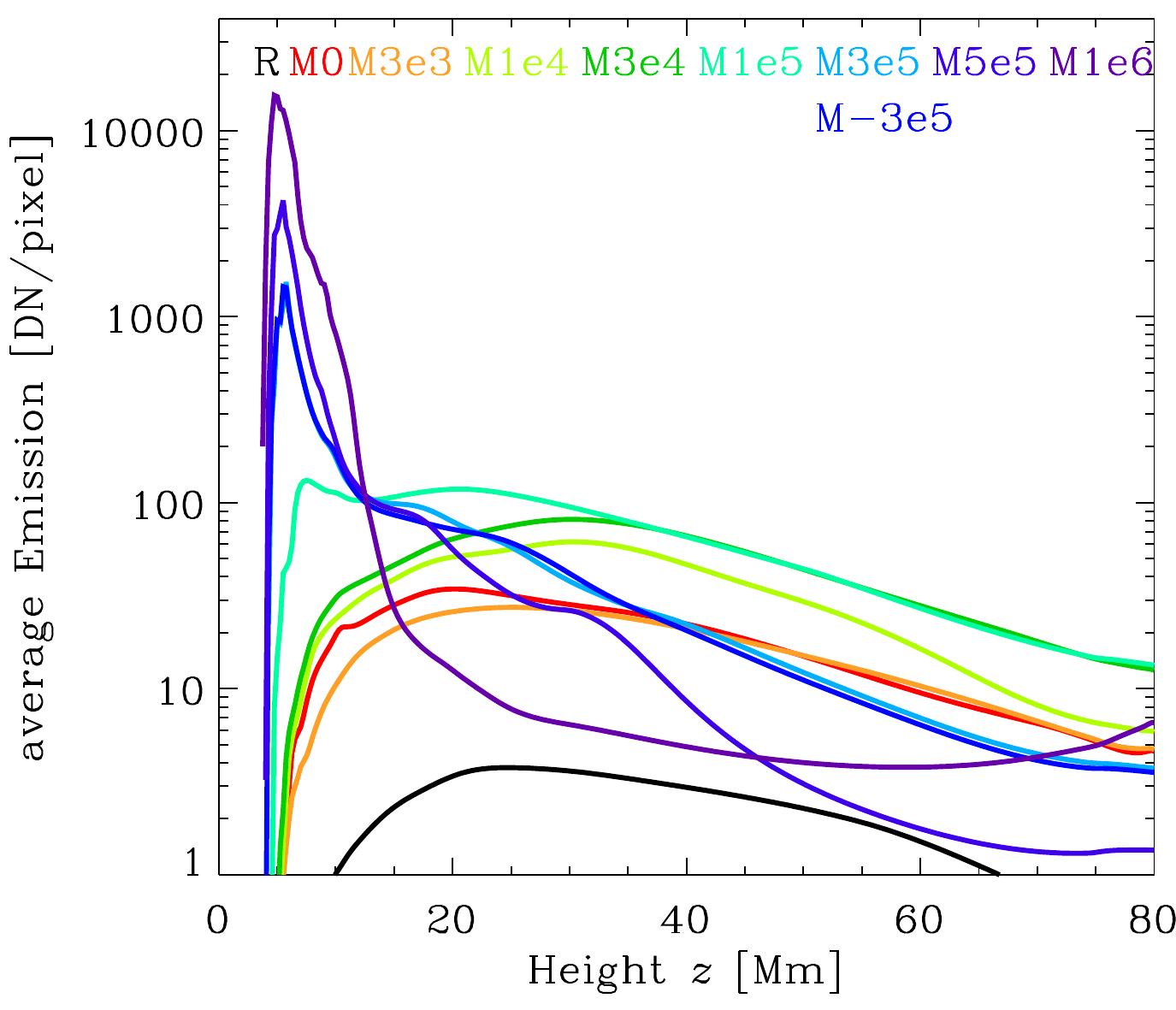}
\end{center}\caption[]{Height dependency of horizontal averaged
  emission of the AIA~171~\AA\ channel. The color of the lines
  indicate the different runs as displayed in the top of the plot.
}\label{emis_height}
\end{figure}

To further show, how the UV emission depends on injected helicity, we
plot in \Fig{emis_height}, the height dependence of the horizontal
averaged UV emission.

\begin{acknowledgements}
The simulations have been carried out on supercomputers at
GWDG, on the Max Planck supercomputer at RZG in Garching and in the facilities hosted by the CSC---IT
Center for Science in Espoo, Finland, which are financed by the
Finnish ministry of education. 
J.W.\ acknowledges funding by the Max-Planck/Princeton Center for
Plasma Physics and 
 from the People Programme (Marie Curie
Actions) of the European Union's Seventh Framework Programme
(FP7/2007-2013) under REA grant agreement No.\ 623609.
\end{acknowledgements}

\bibliographystyle{aa}
\bibliography{paper}

\end{document}